\documentclass[aps,pra,twocolumn,amsfonts,amssymb,amsmath,floatfix,superscriptaddress]{revtex4}

\usepackage{graphicx}
\usepackage{physymb}
\usepackage{color}
\usepackage{subfigure}
\usepackage{bm}
\usepackage{epstopdf}
\usepackage{comment}

\begin{document}
\title{Dynamics of correlations in a quasi-$2$D dipolar Bose gas following a quantum quench}

\author{Stefan S. Natu}
\affiliation{Joint Quantum Institute and Department of Physics, University of Maryland, College Park, Maryland 20742-4111, USA}
\affiliation{Condensed Matter Theory Center, Physics Department, University of Maryland, College Park, Maryland 20742-4111, USA}

\author{L. Campanello}
\affiliation{Condensed Matter Theory Center, Physics Department, University of Maryland, College Park, Maryland 20742-4111, USA}

\author{S. Das Sarma}
\affiliation{Joint Quantum Institute and Department of Physics, University of Maryland, College Park, Maryland 20742-4111, USA}
\affiliation{Condensed Matter Theory Center, Physics Department, University of Maryland, College Park, Maryland 20742-4111, USA}


\begin{abstract} 
We study the evolution of correlations in a quasi-$2$D dipolar gas driven out-of-equilibrium by a sudden ramp of the interactions. On short timescales, roton-like excitations coherently oscillate in and out of the condensate, giving rise to pronounced features in the time-evolution of the momentum distribution, excited fraction and the density-density correlation function. The evolution of these correlation functions following a quench can thus be used to probe the spectrum of the dipolar gas. We also find that density fluctuations induced by the presence of rotons following the quench, dramatically slows down the rate of spreading of correlations in the system: near the roton instability, correlations take infinitely long to build up and show deviations from light-cone like behavior. 
\end{abstract} 
\maketitle

\section{Introduction}

Recent advances in ultra-cold atomic and molecular gases have greatly expanded the potential of these systems as tools for studying many-body physics \cite{blochreview}. For example, the realization of quantum degenerate gases with large magnetic dipole moments \cite{pfau, pfau2, dysprosium1, dysprosium2, erbium}, ongoing efforts to trap and cool polar molecules \cite{jinpolar, zwierleinpolar, weidemullerpolar, schreckpolar}, experiments on Rydberg atoms \cite{weidemuller}, trapped ions \cite{ion1, ion2}, and atoms in high finesse optical cavities \cite{esslinger} have opened up the possibility of realizing ultra-cold atomic systems with long-range interactions, such as the anisotropic dipole-dipole interaction \cite{lahayereview, carrreview, blatt, cirac}. Concurrently, better control over experimental parameters and high resolution imaging techniques have introduced new probes for exploring many-body physics, notable among which is the ability to study the dynamics of correlations following a non-adiabatic ramp (quench) of system parameters \cite{chengstruc, chengresolved, blochresolved, greinerresolved, blochlightcone, chengsakh, blochfermiexp, ddcorrbloch, jinquench}. Remarkably, these quench experiments provide a wealth of information about the low energy properties of the underlying system such as the nature of the excitations, and the manner in which correlations develop in a system. Here we study the evolution of correlations in a  quasi-$2$D dipolar gas following a sudden quench of the interaction strength, finding non-trivial dynamics even in this weakly interacting system. 

\begin{figure}[htr]
\begin{picture}(150, 100)
\put(-15, -5){\includegraphics[scale=0.45]{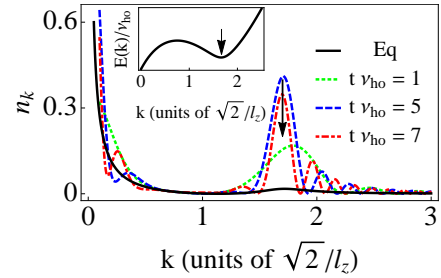}}
\end{picture}
\caption{\label{nexkquench}(Color Online) Top: Evolution of the radial momentum distribution $n_{\textbf{k}}$ in a quasi-$2$D dipolar gas ($l_{z}$ is the width of the cloud in the direction perpendicular to the $2$D plane), following a sudden ramp of the dipolar interaction. Momentum distribution develops a peak at the wave-vector (black arrow) corresponding to the roton minimum in the inset. The period of the oscillations is proportional to the roton gap. Black curve is the the corresponding \textit{equilibrium} feature for the same interaction where no prominent roton signature is found. Time ($t$) is given in units of $1/(\nu_{\text{ho}} = 2\hbar^{2}/ml^{2}_{z})$.} 
\end{figure}

A novel property of quasi-$2$D dipolar superfluids is that in addition to the phonon-like mode at low energies $E_{k} \sim ck$, the low energy excitation spectrum also features a roton-like mode ($E_{k} = \Delta + \hbar^{2}(k-k_\text{r})^{2}/2m^{*}$) for sufficiently strong dipolar interactions \cite{roton, santos}. The existence of roton-like excitations generally implies that the system has a tendency towards developing crystalline order, and the softening of the roton mode is often a route towards realizing correlated states of matter such as supersolids and Wigner crystals in Bose and Fermi systems  \cite{helium,cherng,vengalattore,girvinroton, pohl, sarmass, roscilde}. Observing this mode therefore constitutes an essential first step towards the study of many body physics in dipolar systems. 

Here we investigate how the roton mode is revealed in the spatio-temporal evolution of one and two-body correlation functions, following a sudden switching on of the dipolar interactions. Our main result, summarized in Fig.~\ref{nexkquench}, shows that these correlations develop striking features that are directly related to the underlying roton minimum in the dispersion. This is because on timescales short compared to the collision time, excitations are coherent, and the momentum distribution develops oscillations at different $k$, at a frequency equal to the energy of the excitation $E_{k}$. As the roton excitations oscillate at nearly the same energy ($E_{k} \sim \Delta$), a cooperative effect occurs whereby the amplitude of oscillations is enhanced. The momentum distribution, which is readily measured following time-of-flight, can therefore probe the roton gap.

In addition to providing information about collective excitations \cite{demler1d}, quench experiments raise fundamental questions about the non-equilibrium dynamics of isolated quantum systems \cite{blochlightcone, chengsakh, polkreview}. One such question concerns the manner in which correlations build up between initially uncorrelated regions on short and long times following a quench. For lattice systems interacting via short range forces, Lieb and Robinson showed that correlations evolve in a light-cone manner \cite{blochlightcone, liebrobinson, cardy, hastings}, which has been recently confirmed both experimentally and numerically \cite{blochlightcone}. However, questions remain about continuum and lattice systems interacting with \textit{long range} interactions.  Recent work on trapped ion chains have provided evidence for the breakdown of the light-cone picture in systems with long range interactions \cite{monroelightcone, blattlightcone, daley}. We show that for the continuum system, for weak dipolar interactions, correlations exhibit some features analogous to a light-cone. However the dynamics show marked deviations from light-cone evolution for strong dipolar interactions: correlations take longer to build up, and one cannot associate a characteristic velocity with their spread. This is due to the oscillation of the system between a weakly interacting superfluid, and a correlated state with locally ordered domains. 

This paper is organized as follows: In Sec II we present the time-dependent Bogoliubov approach for studying the evolution of correlations in weakly interacting quasi-$2$D dipolar gases and discuss our results in Sec III. In Sec IV, we compare our results with those of a fully self-consistent theory, recently developed by Lin and Radzihovsky \cite{radzihovsky}, where the condensate fraction is considered as a time-dependent variable. In Sec V, we discuss the relevance of our work to ongoing experiments on dipolar Bose gases, and summarize our results in Sec VI.


\section{Theory}

We consider a quasi-$2$D dipolar Bose gas of mass $m$, at zero temperature, confined in a harmonic potential of the form $U(z) = \frac{1}{2}m\omega^{2}_{z}z^{2}$, and free in the $x-y$ (transverse) directions. Furthermore, we assume that the dipoles are polarized along the $z-$direction, which yields a dipolar interaction potential of the form $V_{\text{dip}}(\textbf{R}) = \frac{d^{2}}{|\textbf{R}|^{3}}(1 - 3\cos^{2}(\theta))$, where $d$ is the dipole moment, and $\cos (\theta) = z/|\textbf{R}|$. Additionally, there is a short-range interaction potential, which we model as $V_{\text{c}}(\textbf{R}) = g\delta(\textbf{R})$, where $g = 4\pi\hbar^{2}a/m$, and $a$ is the $3$D s-wave scattering length. Effects of finite temperature ($T$) are small provided $T \ll \Delta$ \cite{boudgemaa}. 

We model the quasi-$2$D gas by making the Gaussian ansatz: $n^{\text{3D}}(\textbf{R} = \{\textbf{r}, z\}) = n(\textbf{r})\Psi(z) = \frac{1}{\sqrt{\pi l^{2}_{z}}}n(\textbf{r})e^{-z^{2}/l^{2}_{z}}$, where $\textbf{r}$ and $z$ is the radial and the axial co-ordinate respectively, and $l_{z}$ is a variational parameter which can be determined by solving the Gross-Pitaevskii equation in the axial direction \cite{fischer}. When the confining potential greatly exceeds the interaction energy, $l_{z} \rightarrow  \sqrt{\hbar/m \omega_{z}}$, while for weak axial confinement, $l_{z} \sim \zeta$, the healing length of the condensate. Integrating out the $z$-direction, the Fourier transform of the resulting quasi-$2$D interaction potential reads \cite{fischer, wilsonsf}:
$V(k) = \frac{1}{\sqrt{2\pi}l_{z}}\Big(g + g_{d}~F\Big(\frac{{k}l_{z}}{\sqrt{2}}\Big)\Big)$,
where $k = \sqrt{k^{2}_{x}+k^{2}_{y}}$ is the magnitude of the radial momentum, $g_{d} = \frac{2\pi}{3}d^{2}$, and $F(x) = 2 - 3\sqrt{\pi}x~\text{Erfc}(x)e^{x^{2}}$, where $\text{Erfc}(x)$ is the complimentary error function. 


The Hamiltonian for a uniform quasi-$2$D dipolar Bose gas, where the dipoles are aligned along the $z-$axis reads:
\begin{eqnarray}\label{boseham}
{\cal{H}} = \sum_{\textbf{k}}\left(\epsilon_{k}- \mu\right)a^{\dagger}_{\textbf{k}}a_{\textbf{k}}  + \frac{1}{2\Omega}\sum_{\textbf{p} \textbf{q} \textbf{k}}V(q)a^{\dagger}_{\textbf{p}+\textbf{q}}a^{\dagger}_{\textbf{k}-\textbf{q}}a_{\textbf{k}}a_{\textbf{p}}
\end{eqnarray}
where $\epsilon_{k} = \hbar^{2}k^{2}/2m$, $a_{\textbf{k}}$ is the bosonic annhilation operator at momentum $\textbf{k}$ and time $t$, $\Omega$ is the area, and $\mu$ is the chemical potential. 


We consider the evolution of the momentum distribution $n_{\textbf{k}}(t) = \langle a^{\dagger}_{\textbf{k}}(t)a_{\textbf{k}}(t) \rangle$, and the density-density correlation function: 
$g^{(2)}(\textbf{r}, t) = \sum_{\textbf{q}}e^{i \textbf{q}\cdotp\textbf{r}}\langle \rho_{\textbf{q}}(t)\rho_{-\textbf{q}}(0) \rangle$, where $\rho_{\textbf{q}}(t) = \sum_{\textbf{k}}a^{\dagger}_{\textbf{k}+\textbf{q}}(t)a_{\textbf{k}}(t)$, following a sudden quench in the dimensionless interaction parameter $\tilde g =  g_{d}/g$. The former can be readily probed in time-of-flight \cite{blochreview2}, while the latter can be studied using high resolution imaging \cite{blochlightcone, chengsakh}, Bragg spectroscopy \cite{braggspectroscopy} or noise correlations \cite{noisecorrelations}. 

\subsection{Time-dependent Bogoliubov ansatz}

In quasi-$2$D at zero temperature, there is a true Bose condensate \cite{petrov}, and we can model the dynamics using a time-dependent Bogoliubov approach. This theory is valid for weak interactions, which are governed by the small parameter $a/l_{z} \ll 1$. Throughout we refer to ``weak" dipolar interactions as the regime where the underlying dispersion does not have a roton feature and ``strong" dipolar interactions as the regime where it does. It is important to emphasize that the physics of the roton mode in quasi-$2$D dipolar gases is \textit{not} a feature of strong density correlations (as in He-$4$ \cite{griffinbook}), rather one of geometric confinement \cite{santos}. Hence the condition $a/l_{z} \ll1$ is satisfied even for ``strong'' dipolar interactions, where the roton mode is present. 

We set the density of condensate atoms $n_{0} = \langle a_{\textbf{k} = 0} \rangle^{2}$, and write $a_{\textbf{k} \neq 0}(t) = u_{k}(t)b_{\textbf{k}} + v^{*}_{k}(t)b^{\dagger}_{-\textbf{k}}$, where $b_{\textbf{k}}$ denotes the bosonic annihilation operator for the non-condensed atoms \cite{natuddcorr}. The $b_{\textbf{k}}$ operators have no time dependence, and are formally treated as small. Substituting the expression for $a_{\textbf{k} \neq 0}$ into Eq.~\ref{boseham}, and discarding all terms cubic or higher order in $b_{\textbf{k}}$, we arrive at 
$u_{k}(t = 0) = \sqrt{\frac{1}{2}\left(1 + \frac{\epsilon_{k} + V_{i}(k)n_{0}}{E^{i}_{k}}\right)} $ and $v_{k}(t = 0) = -\text{sgn}(V_{i}(k))\sqrt{\frac{1}{2}\left(\frac{\epsilon_{k} + V_{i}(k)n_{0}}{E^{i}_{k}} - 1\right)}$, 
where $\text{sgn}(x)$ denotes the sign of the argument, and $E^{i}_{k} = \sqrt{\epsilon_{k}(\epsilon_{k} + 2 V_{i}(k)n_{0})}$, where the index $i$, denotes the initial state. At future times, these coherence factors $u_{k}(t)$ and $v_{k}(t)$ acquire complex values, but will always satisfy $|u_{k}(t)|^{2} -|v_{k}(t)|^{2} =1$. 

For now we assume that the condensate is static, in other words, $n_{0} \approx n$. The advantage of this somewhat simplified approximation is that it allows us to make analytic predictions for the long time behavior of the excited fraction. Later on we present a fully self-consistent Bogoliubov theory which takes into account the time-dependence of the condensate density via the relation $n = n_{0}(t) + \sum_\textbf{k}n_{\textbf{k}}$, where $n$ is the total density, which is a constant of motion. This allows us to put better quantitative bounds on the validity of our theory for experiments on quasi-$2$D dipoles.

The equations of motion for $u_{k}$ and $v_{k}$ are obtained from the Heisenberg equations of motion for $a_{k}$, and read \cite{natuddcorr}:
\begin{eqnarray}\label{eom}
i\partial_{t}\left(\begin{array}{cc} 
u_{k}(t) \\
v_{k}(t) \end{array}\right) =\left(\begin{array}{cc} A_{k} & B_{k} \\ -B_{k} & -A_{k}\end{array}\right)
\left(\begin{array}{cc} 
u_{k}(0) \\ 
v_{k}(0) \end{array}\right)
\end{eqnarray}
where we have introduced the functions $A_{k} = \epsilon_{k} + V_{f}(k) n_{0}$, $B_{k} = V_{f}(k) n_{0} $ and $E^{f}_{k} = \sqrt{A^{2}_{k}- B^{2}_{k}}$ is the Bogoliubov dispersion. As we consider a sudden quench, the evolution of $u_{k}$ and $v_{k}$ depends only on the final Hamiltonian parameters, denoted by the label $f$. Physically, the time-dependent Bogoliubov approximation amounts to coherent oscillations of quasi-particles in and out of the condensate at a frequency proportional to the quasi-particle energy $E^{f}_{k}$. Absent collisions, excitations at different momenta evolve independently of one another. We note that Eq.~\ref{eom} is the bosonic analog of the time-dependent Bogoliubov deGennes equations, which describe the dynamics of Cooper pairs following an interaction quench in a Fermi gas \cite{levitov}.

The evolution of the momentum distribution is given by $n_{\textbf{k}}(t) = |v_{k}(t)|^{2}$. We also define the excited fraction as $n_{\text{ex}}(t) = \int d\textbf{k} ~n_{\textbf{k}}(t)$. The density-density correlation function takes the form $g^{(2)}(\textbf{r}, t) = n^{2}_{0} + n_{0}\sum_{\bf{k}}e^{i \bf{k}\cdotp\textbf{r}}\Bigl(2|v_{k}(t)|^{2} + u^{*}_{k}(t)v_{k}(t) + u_{k}(t)v^{*}_{k}(t) \Bigr)$ \cite{glauber}. The first term is the correlation between the condensate atoms, while the second term involves correlations between the condensed and non-condensed atoms. Terms which involve correlations between the non-condensed atoms alone (quartic in $u_{k}$s and $v_{k}$s) are negligible on length scales $\textbf{r} \gtrsim \zeta$ \cite{stoof, natuddcorr, tan}.  We define a dimensionless density-density correlation function $\tilde g^{(2)}(t) = g^{(2)}(t)l^{2}_{z}/2n_{0}$.

\begin{figure}[ht]
\begin{picture}(200, 170)
\put(15, 67){\includegraphics[scale=0.42]{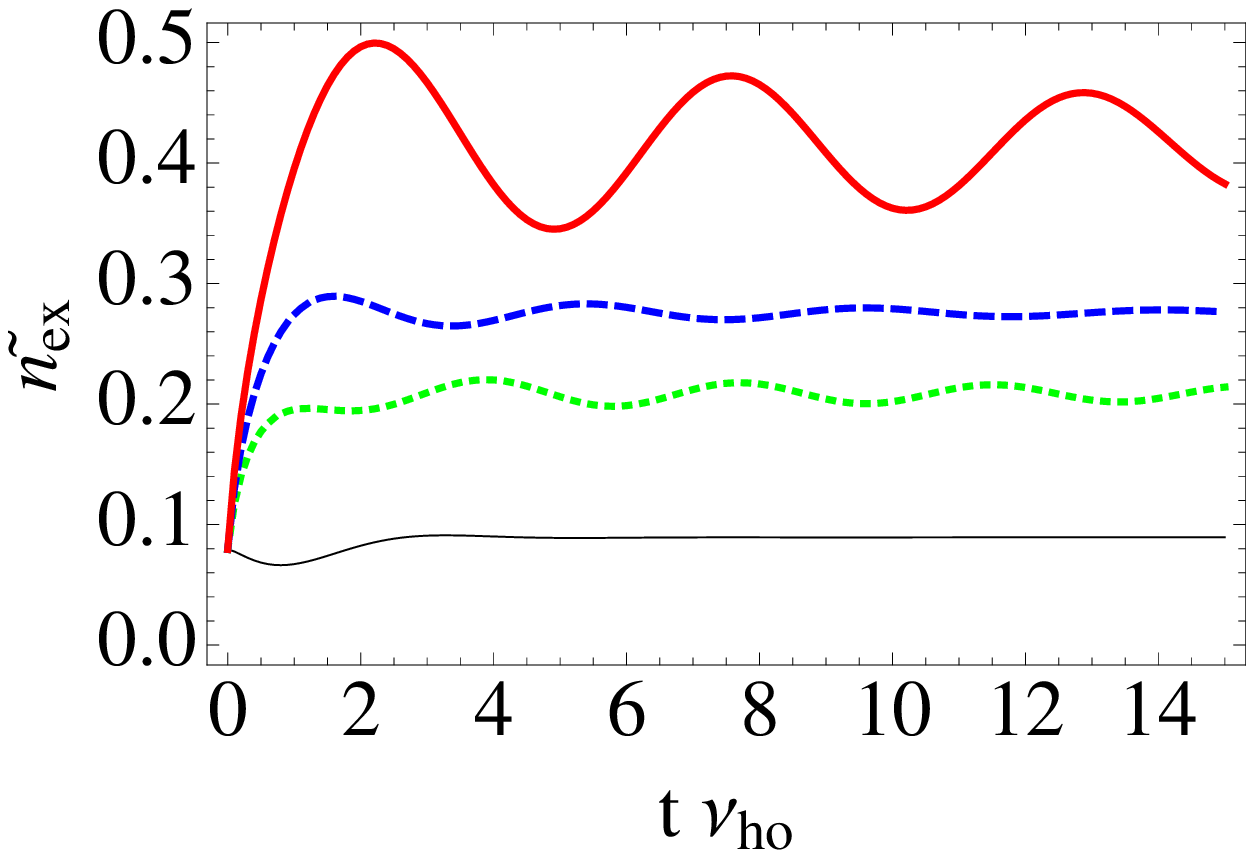}}
\put(-20, -10){\includegraphics[scale=0.32]{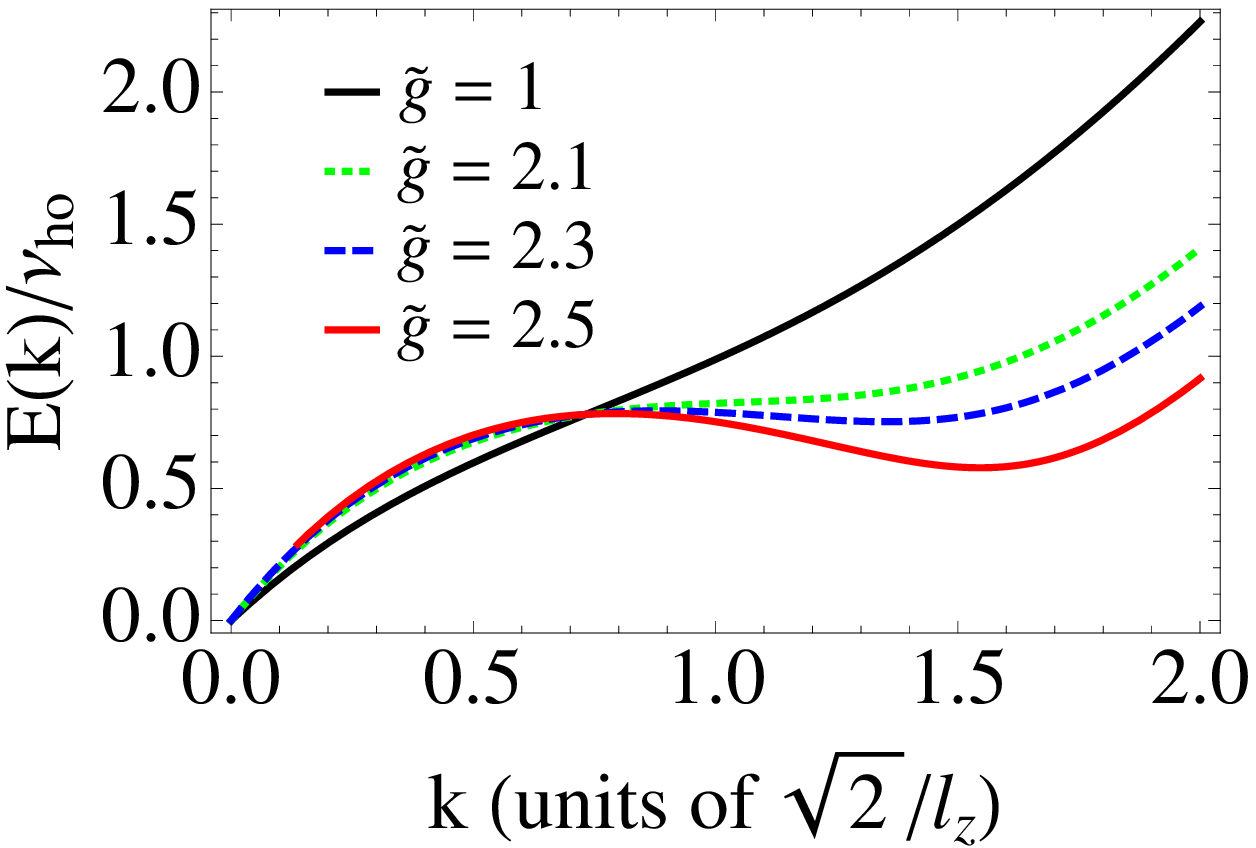}}
\put(103, -10){\includegraphics[scale=0.32]{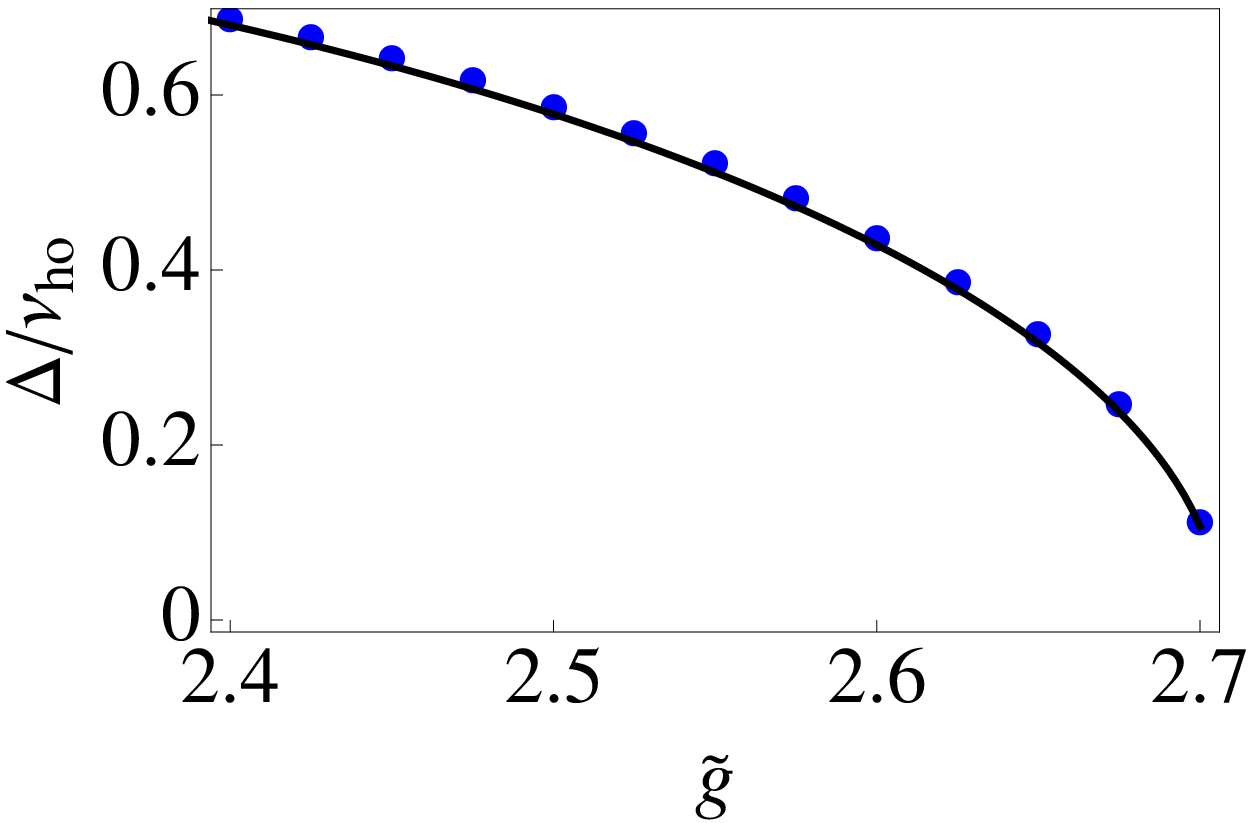}}
\end{picture}
\caption{\label{nextplot} (Color Online) Top: Coherent evolution of the dimensionless excited fraction $\tilde n_{\text{ex}}(t) = l^{2}_{z}n_{\text{ex}}/2$ for different values of $\tilde g_{f} = g_{d}/g$. To compare the scale on the vertical axis, the total density is $nl_{z}^{2} \sim 3$ for typical densities and axial trapping potentials. From bottom to top: $\tilde g_{f} = 1$ (thin black), $2.1$ (dotted green), $2.3$ (dashed blue), $2.5$ (thick red). Bottom-left: energy-momentum dispersion curves for the values of $\tilde g_{f}$ in Top. Bottom-Right: Data shows numerically extracted values of the roton gap from the oscillations in the top plot (see solid red curve), solid line is the roton gap extracted from the energy momentum dispersion. For $\tilde g < 2.3$ the dispersion does not have a sharp roton-like feature.}
\end{figure}

\section{Results}

We now present the results for the evolution of the momentum distribution and the density-density correlation function following a quench using the non-self-consistent version of the Bogoliubov theory presented above. In the next section, we will go beyond this approximation by making the condensate density time-dependent.

Although we can model arbitrary quenches, throughout we consider an initial state prepared in equilibrium at zero temperature, with $\tilde c^{2} = g_{i}n_{0}/\nu_{\text{ho}} = 1$, where $\nu_{\text{ho}} = 2\hbar^{2}/ml^{2}_{z}$ and $\tilde g_{i} = g_{d}/g = 0$. The relevant parameter  governing the dynamics is the ratio of the strength of the dipolar and contact interactions after the quench $\tilde g_{f} = g_{d,f}/g_{f}$. Experimentally this parameter can be varied by either tuning the contact interaction or the magnetic dipole-dipole interaction via a rapidly varying planar magnetic field in addition to the static field along $z$ \cite{pfau, dipoletune}. As discussed by Giovanazzi \textit{et al.} \cite{dipoletune}, this latter protocol yields an effective (time-averaged) dipolar interaction of the form $\langle V_{\text{dip}}(\textbf{R}) \rangle = \frac{d^{2}}{|\textbf{R}|^{3}}(1 - 3\cos^{2}(\theta))\alpha(\phi)$, where $\alpha(\phi) = (3 \cos^{2}(\phi) -1)/2$, where $\phi$ is the tilting angle. By controlling the angle $\phi$, the dipolar interaction can be completely switched off and turned on. Henceforth, we keep $g$ fixed ($g_{i} = g_{f}$), and produce a non-zero $\tilde g_{f}$ by suddenly switching on the dipolar interaction $g_{d}$, and study the subsequent evolution of correlations. The final value of $g_{f}$ is chosen such that the quasi-$2$D dipolar gas is mechanically stable ($E^{f}_{k} \geq 0$ for all $k$) \cite{fischer}.

In Fig.~\ref{nexkquench}, we plot the temporal evolution of the momentum distribution $n_{\textbf{k}}(t)$ for a quench to strong dipolar interactions ($\tilde g_{f} = 2.65$). In addition to the usual divergence near $k=0$, which is associated with Bose condensation, the momentum distribution develops a second peak associated with roton-like excitations. Furthermore (and somewhat surprisingly), at the peak maximum, the occupation of rotons is substantially larger than the \textit{equilibrium} (zero temperature) occupation of these modes at the same interaction strength. 

In Fig.~\ref{nextplot}(top) we plot the excited fraction obtained by integrating the momentum distribution over all $k$. For weak dipolar interactions, the excited fraction rapidly saturates to a constant value whereas for stronger dipolar interactions, it develops oscillations which become pronounced as $\tilde g_{f}$ is increased. 

Physically, this is understood as follows: after the quench, the system responds by populating modes at different wave-vectors. For a non-dipolar gas, as quasi-particles oscillate in and out of the condensate, ``fast" quasi-particles rapidly dephase relative to one another, leaving only the ``slow" modes, namely the \textit{phonons}. The energy scale separating the slow and fast modes is $(g+ g_{d})n_{0}$, and consequently, the excited fraction saturates to its asymptotic value on times $\tau = 1/(g+ g_{d})n_{0}$. 

As the dipolar interaction strength is increased (green, dotted curve in Fig~\ref{nextplot} (top)), the excited fraction develops undamped oscillations. This is due to the fact that for this interaction strength, the dispersion develops a broad flat shoulder (see green (dotted) dispersion curve). As a result, many modes oscillate at the same frequency, again giving rise to undamped oscillations in $n_{\text{ex}}$. This broad shoulder is the precursor to the roton minimum which appears for larger interaction strengths. 

Upon further increasing the dipolar interaction, (equivalently $\tilde g_{f}$) the excited fraction shows pronounced, weakly damped oscillations. This is because for strong dipolar interactions, there are two distinct ``slow" modes in the system: phonons and rotons. These modes have different dispersions, hence different density of states (DOS): the phonon DOS $n(\epsilon) = k~dk/d\epsilon$ vanishes as $\epsilon \rightarrow 0$, whereas the roton DOS reads $n(\epsilon)  =  m^{*}(1 + k_{\text{r}}/\sqrt{2m^{*}|\epsilon - \Delta|)}$, which diverges as $1/\sqrt{|\epsilon-\Delta|}$. For quenches to strong dipolar interactions, the larger DOS for roton-like modes implies that the system preferentially occupies rotons following the quench. Furthermore, unlike phonon modes which occupy a range of momenta ($0 < \hbar k < \sqrt{2m(g+g_{d})n_{0}}$), all the roton modes have nearly the same energy ($E_{\text{roton}} \sim \Delta$). This leads to a cooperative amplification of the oscillations in the momentum distribution. Not surprisingly, the timescale for these oscillations is roughly $\tau_{\text{roton}} = 1/\Delta$. For quenches to strong dipolar interactions, where $\tau_{\text{roton}} > \tau$, the timescale for the fast modes to dephase relative to one another, the excited fraction also develops oscillations. These can be readily probed in time-of-flight \cite{jinquench}. The dynamics of the roton modes in this \textit{bosonic} dipolar gas is analogous to that of Cooper pairs in an attractive Fermi gas, driven out of equilibrium following an interaction quench \cite{levitov}. In that case, the coherent oscillation of Cooper pairs leads to oscillations in the superconducting pairing gap.

At long times, the excited fraction calculated using a saddle point approximation takes the approximate analytic form $n_{\text{ex}}(t \rightarrow \infty) \propto \cos(2\Delta t + \phi)/\tilde\Delta^{2}\sqrt{\eta~t~\nu_{\text{ho}}}$, where $\eta = m/m^{*}$, $\tilde \Delta = \Delta/\nu_{\text{ho}}$, and $\phi$ is an arbitrary constant phase factor. The dynamics are described by damped oscillations at a frequency equal to twice the roton gap, and an envelope which decays algebraically as $1/\sqrt{t}$. Indeed, the roton gap extracted from fitting the numerical data for $n_{\text{ex}}(t)$ to this long time asymptotic formula is in near perfect agreement with the gap obtained directly from the dispersion relation $E_{k} = \sqrt{\epsilon_{k}(\epsilon_{k} + 2 V_{f}(k)n_{0})}$. Hence the roton gap can be directly accessed from a sudden quench experiment. 



\begin{figure}[htr]
\begin{picture}(200, 210)
\put(-20, 70){\includegraphics[scale=0.32]{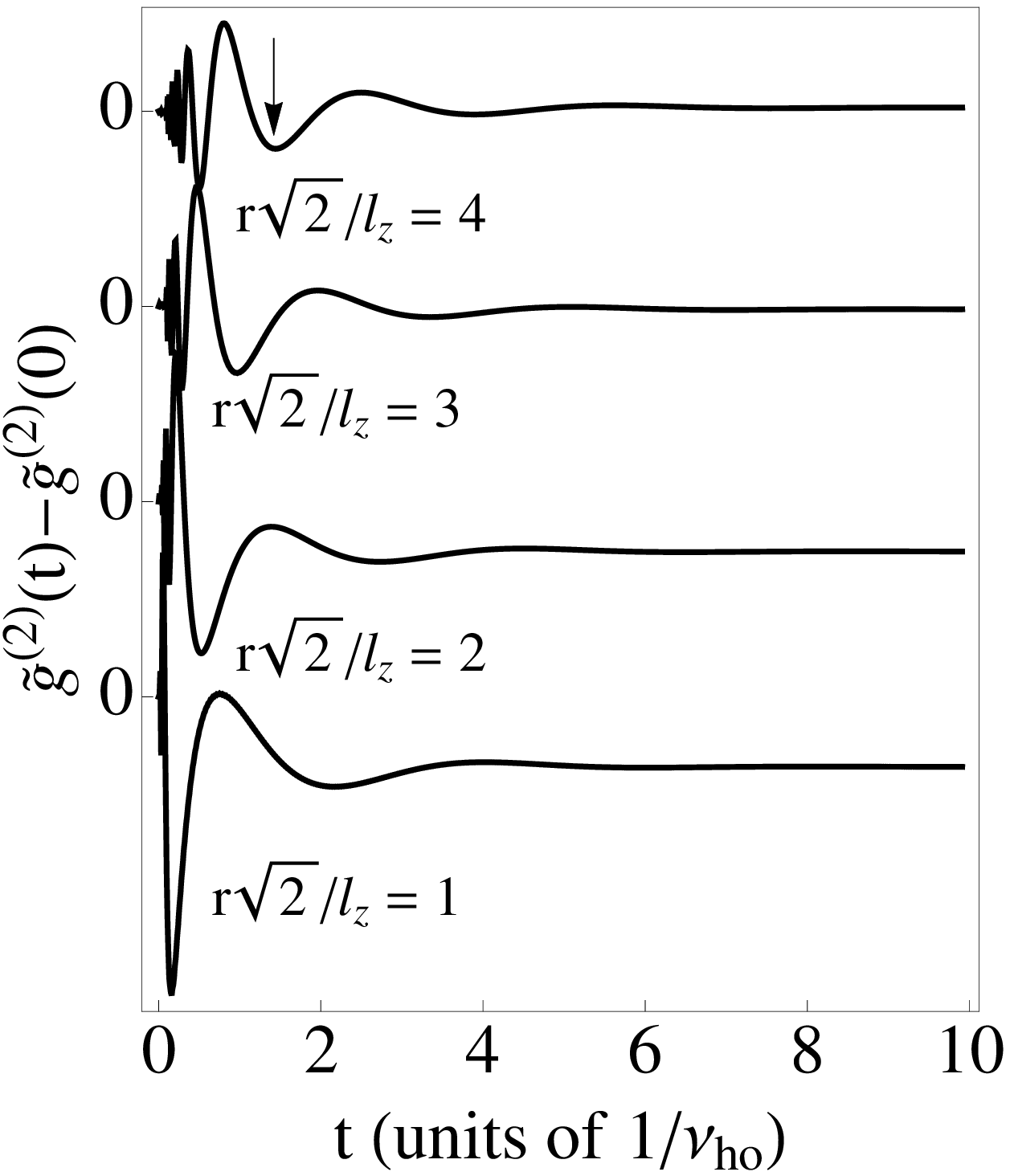}}
\put(105, 70){\includegraphics[scale=0.32]{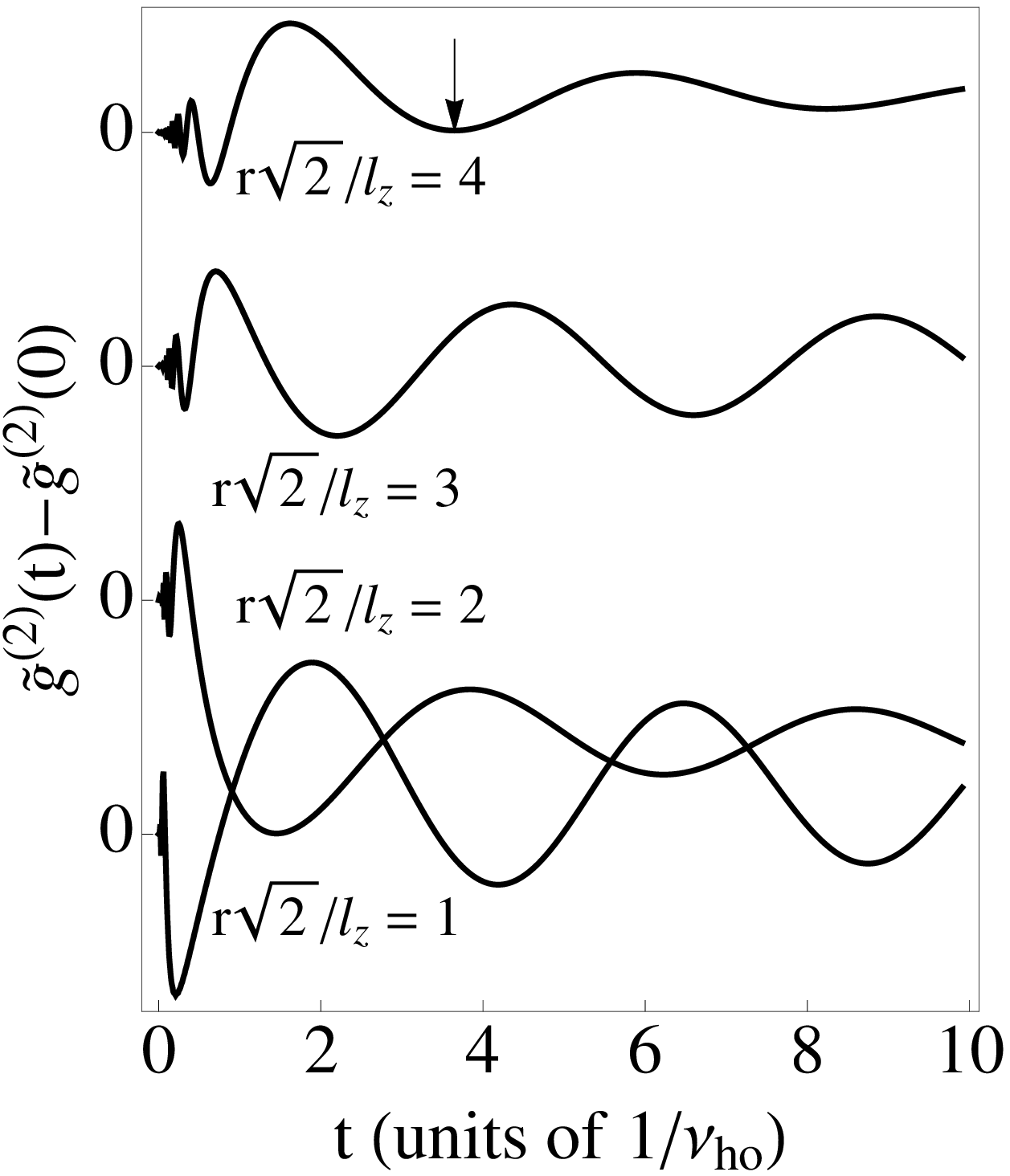}}
\put(-20, -10){\includegraphics[scale=0.3]{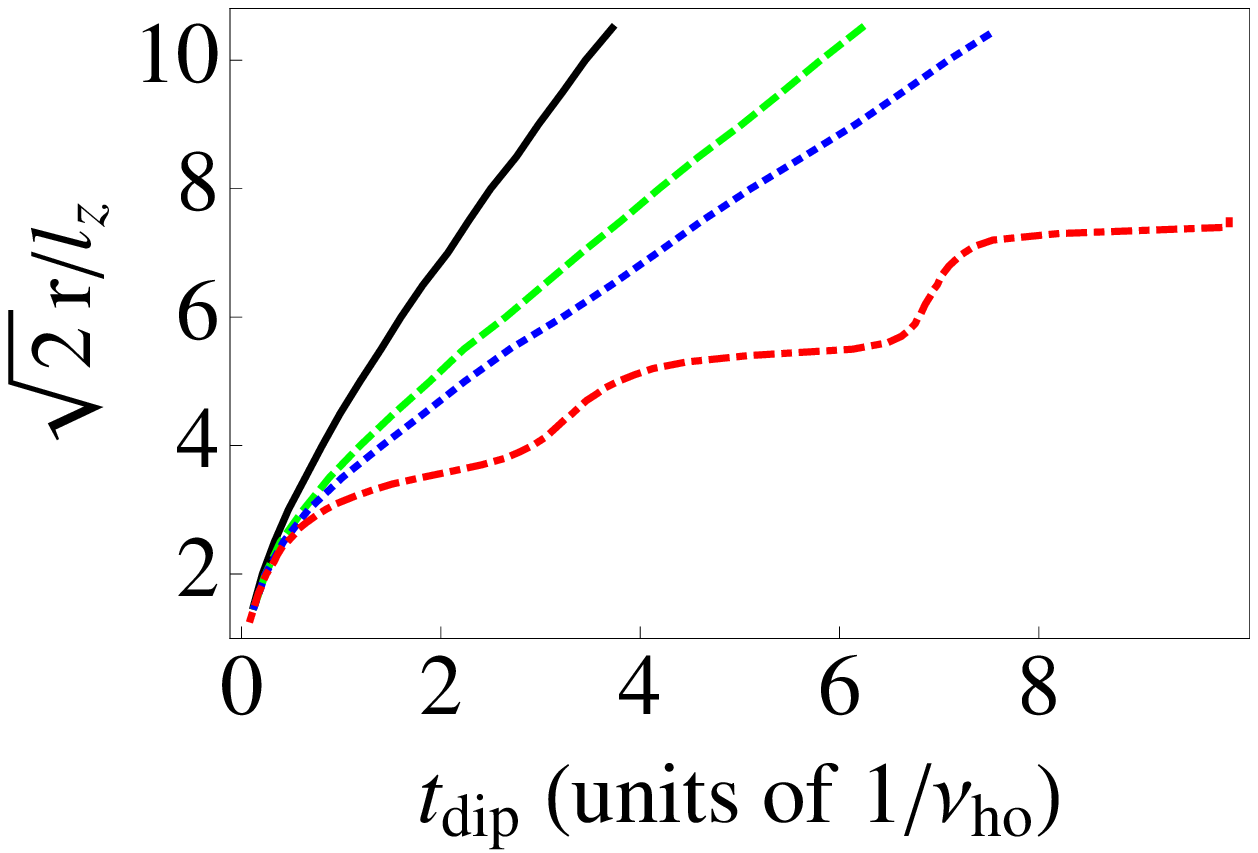}}
\put(100, -10){\includegraphics[scale=0.3]{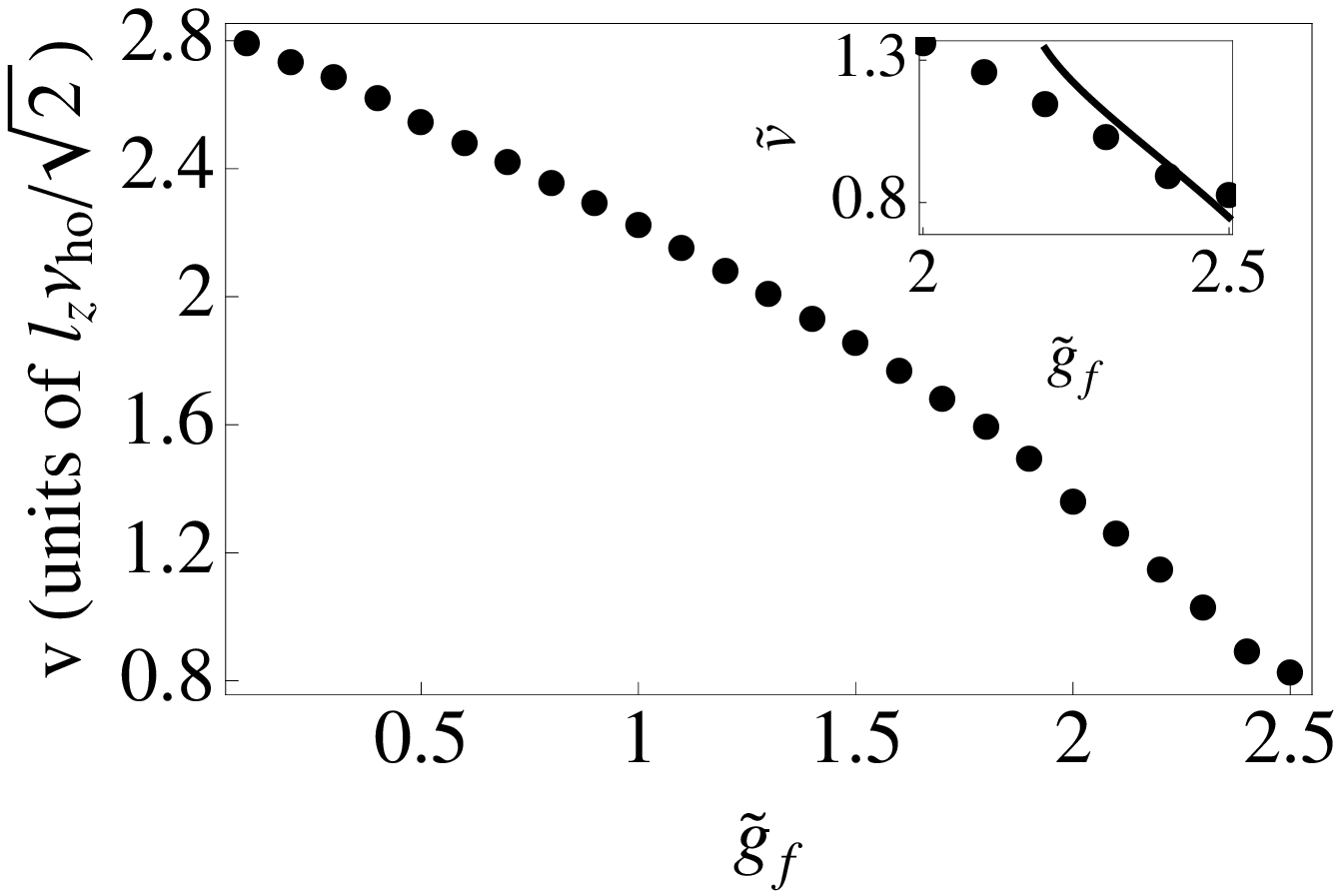}}
\end{picture}
\caption{\label{ddcplot} Top: Density-density correlation function for different values of $\textbf{r}$, for a quench to $\tilde g_{f} = 1$ (left) and $\tilde g_{f} = 2.6$ (right) respectively. Arrows indicate the feature whose dispersion we study in the bottom plot for different final values of $\tilde g_{f}$. (Bottom-Left) Temporal location of the dip feature (arrow) in the density-density correlation ($t_{\text{dip}}$) for different values of $\textbf{r}$ and different interactions: Black $\tilde g_{f} = 1$, Green (dashed) $\tilde g_{f} = 2$, Blue (dotted) $\tilde g_{f} = 2.2$, Red (dashed-dotted) $\tilde g_{f} = 2.6$. (Bottom-Right) Data points show the velocity of spreading of correlations in a quasi-$2$D gas for quenches to different $\tilde g_{f}$. Inset is a zoom-in for strong dipolar interactions ($\tilde v = v/(l_{z}\nu_{\text{ho}}/\sqrt{2})$), the solid line is the characteristic velocity of roton-like modes $v = 2\Delta/\hbar k_{\text{r}}$ for different $\tilde g_{f}$.}
\end{figure}

We now turn to the evolution of the density-density correlation function\cite{blochlightcone, chengsakh, monroelightcone, blattlightcone}, which reads: $\tilde g^{(2)}(t) -\tilde g^{(2)}(0) =  -\frac{1}{\pi}\int dk k J_{0}(kr) \sin(E^{f}_{k} t)^{2} \epsilon_{k} B^{f}_{k}\frac{\epsilon_{k} + g_{i}n_{0}}{E^{i}_{k}E^{f}_{k}}$, where $J_{0}$ is the Bessel function of the first kind. 


In Fig.~\ref{ddcplot} we plot $g^{(2)}(\textbf{r}, t)$ for different values of $\textbf{r}$ following a quench to two different values of $\tilde g_{f}$, starting from $\tilde g = 0$. For weak dipolar interactions ($\tilde g_{f} \lesssim 1$), following rapid oscillation on very short timescales, the density-density correlation function develops a dip feature which disperses to later times as $\textbf{r}$ increases, and then rapidly relaxes to its long time value \cite{natuddcorr, kollath}. For strong dipolar interactions, density correlations display persistent oscillations, with a frequency largely independent of $\textbf{r}$. The rapid oscillations at very short times stem from the propagation of very fast particles ($E_{k} \sim k^{2}, kl_{z} \gg 1$), produced after the quench, and are absent in systems with a bounded spectrum (arising from a lattice) \cite{natuddcorr, blochlightcone, kollath}. 

On longer timescales, correlations develop a dip feature, associated with low energy excitations. The temporal location of the dip feature ($t_{\text{dip}}$) in the density-density correlations (Fig.~\ref{ddcplot} (bottom-left)) shows a linear dependence with $\textbf{r}$ at small $\tilde g_{f}$. The slope of the curves at large $\textbf{r}$, plotted on the bottom-right panel, is the characteristic velocity of spreading of correlations. Absent damping processes, quasi-particles propagate ballistically, hence correlations display  light-cone behavior at large distances \cite{liebrobinson, cardy, natuddcorr}, which a velocity $v \sim \langle\partial E_{k}/\partial k\rangle$. At small $\tilde g_{f}$, we expand the dispersion as $E_{k} \approx \sqrt{\tilde c^{2}(1 + 2\tilde g_{f})}k$, and approximate the Bessel function $J_{0}(x) \approx \cos (x/\sqrt{2})$, to obtain a velocity $v \approx  2 l_{z}\nu_{\text{ho}}$, which is indeed what we find numerically by taking the slope of the black curve in (Fig.~\ref{ddcplot} (bottom-left)).

Increasing dipolar interactions \textit{decreases} the velocity, as repulsive interactions lower the average group velocity of the quasi-particles in a quasi-$2$D dipolar gas. Nonetheless, for $\tilde g \leq 2.5$, correlations still spread in a  light-cone manner, and one can associate a characteristic velocity with their spread. For strong dipolar interactions, this velocity approaches $v \rightarrow 2\Delta/\hbar k_{\text{r}}$.

For quenches to even stronger dipolar interactions, the dispersion of the dip develops a non-trivial step-like feature. The width of the step grows with $\tilde g_{f}$, becoming infinitely wide at the roton instability threshold, indicating that near the roton instability, correlations take infinitely long to build up. Physically, this can be understood as follows: following the quench to strong dipolar interactions, the system develops locally ordered domains, which are otherwise uncorrelated with one another. These domains prevent correlations from spreading through the system. The timescale over which these domains appear and disappear is linked to the oscillations in the roton occupation number, which is set by $\tau_{\text{roton}}$. Correlations spread rapidly when the domains are absent, and slowly when the domains are present, giving rise to the step-like feature. Near the roton-instability threshold, $\tau_{\text{roton}}$ diverges, and the domains become extremely long lived, preventing correlations from spreading altogether. 





\section{Self-consistent Bogoliubov approach}

In this section we discuss in detail the limitations of the approximation of a static condensate, employed to obtain the results presented in Sec IV. This approach is valid provided that $n_{\text{ex}} \ll n_{0}$ at \textit{all times} during the evolution. While this condition is met for weak interactions or short times, it is dramatically violated very near the roton instability threshold, and a better modeling of the condensate is required. 

In this section we compare our ``static" treatment of the time-dependent Bogoliubov approximation \cite{natuddcorr} of Sec III to a fully self-consistent Bogoliubov treatment of the dynamics, developed by Lin and Radzihovski \cite{radzihovsky}, where we let the condensate density fluctuate in time according to:
\begin{equation}\label{fluccond}
n_{0}(t) = n - n_{\text{ex}}(t).
\end{equation}
Here $n_{0}$ is the condensate fraction, $n$ is the total density, which is a constant of motion, and $n_{\text{ex}} = \int d\textbf{k}~n_{\textbf{k}}(t)$, as defined in the main text. 

\begin{figure}[ht]
\begin{picture}(200, 210)
\put(10, 100){\includegraphics[scale=0.45]{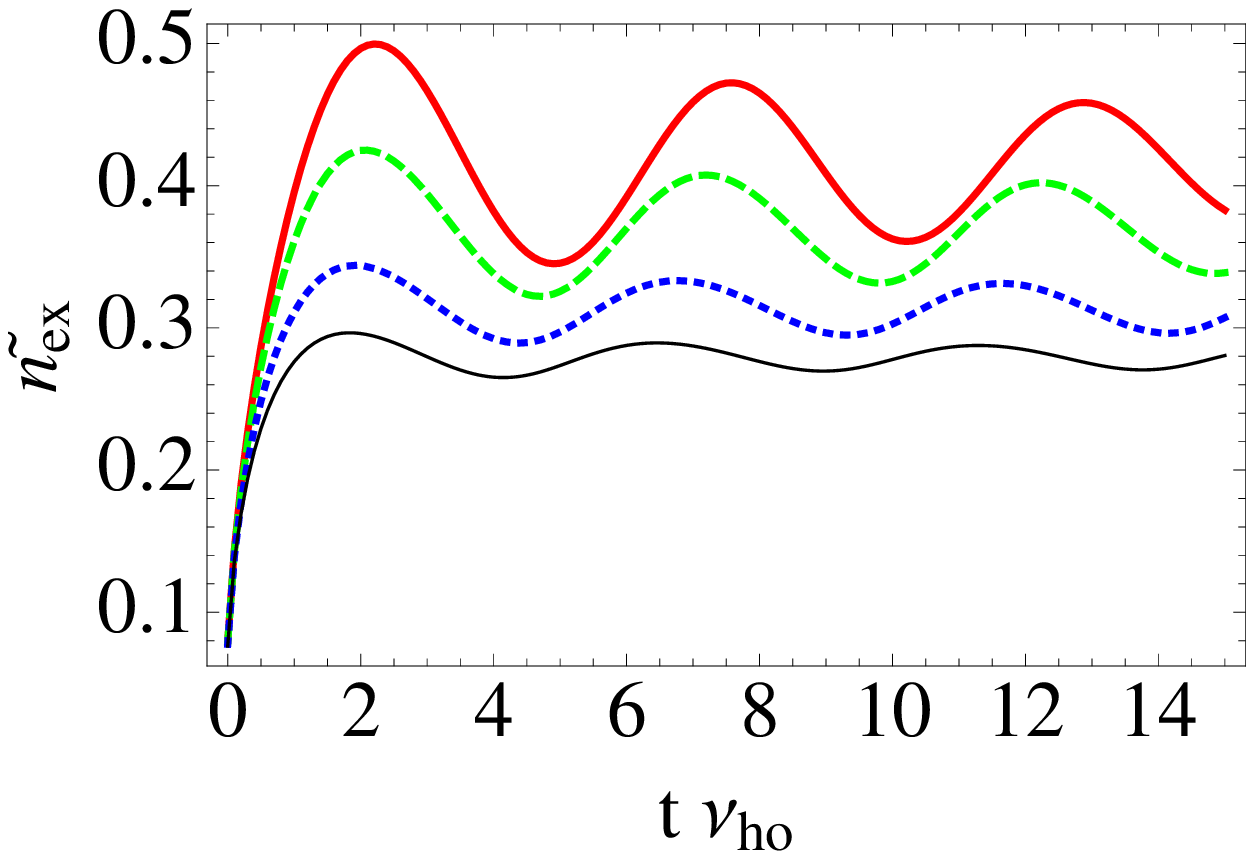}}
\put(10, -10){\includegraphics[scale=0.45]{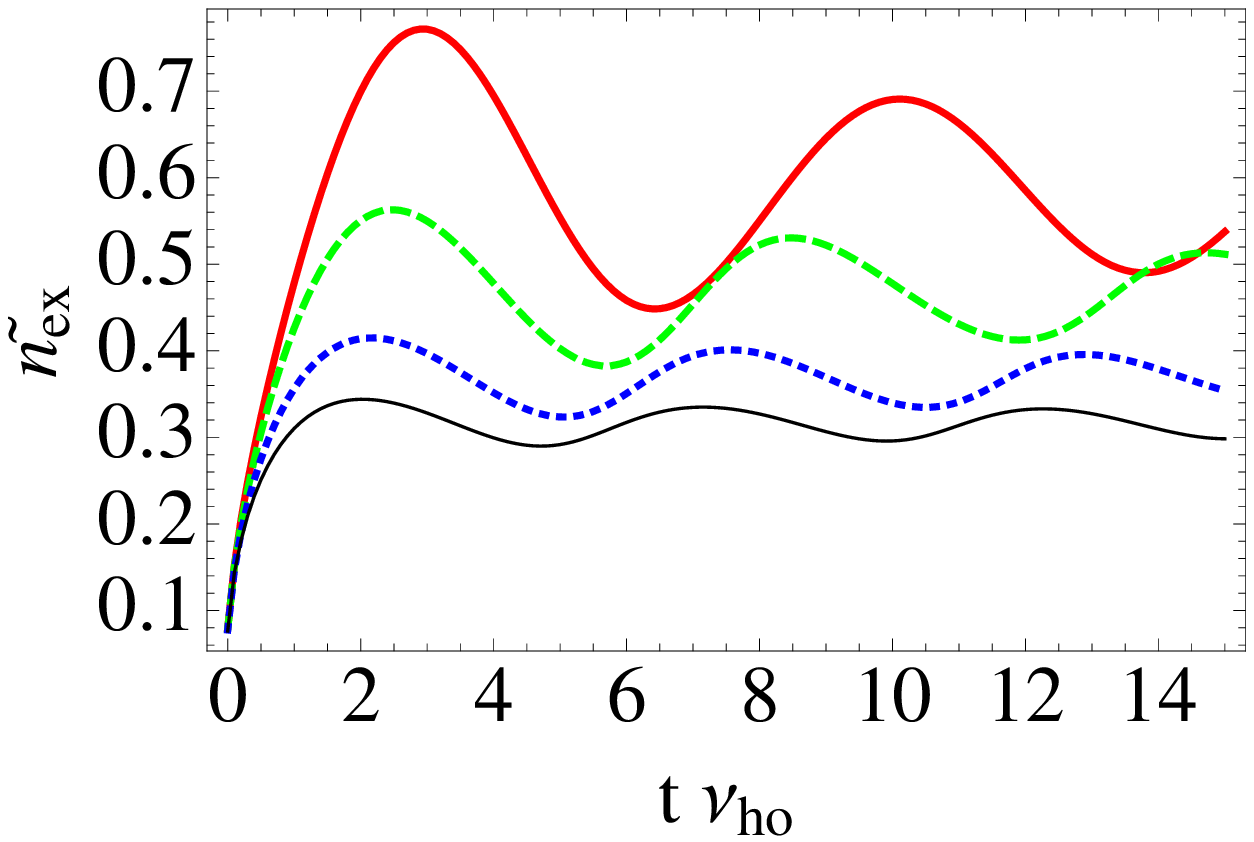}}
\end{picture}
\caption{\label{selfconsistent2} (Color Online) (Top) Coherent evolution of the excited fraction $\tilde n_{\text{ex}}(t)$ for different values of $\eta$ at $\tilde g_{f} = g_{d}/g = 2.5$, obtained by self-consistently solving for the condensate fraction using Eq.~\ref{fluccond}. The red (solid, thick) curve is the non-self consistent result (assuming $n_{0} \approx n$), also shown in Fig.~$2$ of the main text. The green (dashed), blue (dotted) and black (thin) curves correspond to $\eta = 0.05, 0.1, 0.25$ respectively. (Bottom) Same as top with $\tilde g_{f} = 2.6$.}
\end{figure}

As in Sec III, we consider a quench from an initially non-dipolar gas from $\tilde g_{i} = g_{d}/g = 0$, to some finite value $\tilde g_{f} > 0$, keeping the contact interaction parameter $g$ fixed. At each time step ($t$) following the evolution, the condensate fraction is varied according to Eq.~\ref{fluccond}, which is then used to calculate $u_{k}(t+\delta t)$ and $v_{k}(t + \delta t)$ from Eq.~\ref{eom}. The updated excited fraction is used to obtain a new condensate fraction, and the procedure is repeated. This procedure becomes exact in the limit of vanishing $\delta t$ and we have verified that the time-steps ($\delta t$) are small enough that effects of finite step-size are negligible over the timescales we study. 

The relevant parameter that controls the strength of the excited fraction is $\eta = 2/nl^{2}_{z}$, which we vary, along with the quench amplitude $\tilde g_{f}$. Eq.~\ref{fluccond} then becomes:
\begin{equation}\label{fluccond2}
\tilde n_{0}(t) = 1 - \eta~\tilde n_{\text{ex}}(t),
\end{equation}
where $\tilde n_0$ denotes the condensate fraction and $\tilde n_{\text{ex}}$ is the dimensionless excited fraction, which is plotted in Fig.~\ref{selfconsistent2} for a quench to $\tilde g_{f} = 2.5$. Large $\eta$ implies the gas is becoming more $2$D-like. 

Integrating the momentum distribution in Fig.~\ref{selfconsistent2}, we find that the dynamics of the excited fraction is \textit{qualitatively} similar to Fig.~\ref{nextplot}, even when the condensate fraction is assumed to be a dynamical variable. Quantitatively however, the amplitude and frequency of the oscillations decreases as $\eta$ is increased. Although small oscillations are present even for $\eta \gtrsim 0.3$, the amplitude of the oscillations may be difficult to detect experimentally. For the black (thin, solid) curve, the oscillation frequency differs from that shown in the red (thick, solid) curve by $10 \%$. Recall that the oscillation frequency for the red (thick, solid) curve is equal to twice the \textit{equilibrium} roton gap at $\tilde g = 2.5$. Increasing the quench amplitude ($\tilde g_{f}$) causes the oscillations in the excited fraction to become anharmonic, and for larger $\tilde g_{f}$ values, one has to go to smaller $\eta$ to recover quantitative agreement with the non-self consistent results. The anharmonicity is to be expected as the different $k$ modes are now coupled to one another via the time-dependent condensate fraction. Nonetheless, even at $\tilde g = 2.6$, shown in the bottom panel, the dynamics retains the qualitative features obtained from the simple theory presented in the main text. 

The parameter $\eta$ is experimentally tunable either by tuning the density or the axial trapping potential. Based on our findings above, quench dynamics can probe details of the underlying \textit{equilibrium} dispersion, even for strong quenches ($\tilde g \sim 2.5$), provided that $\eta \leq 0.3$. Despite the seemingly small value of this parameter, this implies that the inter-particle spacing $n^{-1/2}/l_{z} \leq 0.4$. These conditions can be readily achieved in experiments even for moderate densities ($n \sim 10^{8} -10^{9}$cm$^{-2}$), for typical trapping potentials of $\omega_{z}/2\pi \sim 100-1000$Hz \cite{boudgemaa}. 


On even longer timescales, beyond mean-field corrections such as collisions between quasi-particles, and between quasi-particles and condensate atoms (Landau and Beliaev damping \cite{stringari}), which are not captured by our self-consistent approach will endow the excitations with a finite lifetime ($\tau_{el}$), and drive the system towards thermal equilibrium. At low temperatures  $T \ll \mu, \Delta$, Beliaev processes dominate, and their rates can be estimated perturbatively in powers of $a/l_{z}$. For quasi-$2$D dipolar gases however, these rates are strongly suppressed for phonons and rotons, owing to the nature of the quasi-$2$D dispersion \cite{landaunatu, beliaevnatu}.

\section{Discussion}

We now discuss the relevance of our work to ongoing experiments on quasi-$2$D dipolar gases.  Experiments so far have not been able to decisively verify the existence of the roton mode, because the signatures of this mode in correlation functions are rather weak \cite{pfauprivate}. Here we have proposed the use of a sudden quench to amplify these signatures. The physical reason for this is clear: on short timescales, excitations coherent oscillate in and out of the condensate and the fact that roton excitations all oscillate with nearly the same frequency will lead to an amplification of the signal, similar to what occurs in driven Fermi systems. In this respect, the bosonic dipolar gas behaves like a fermionic system. The main advantage of our approach is its simplicity; the momentum distribution following a quench can be readily measured in ultra-cold systems, and direct provides access to the roton gap, without the need for sophisticated spectroscopy.

Using a time-depedent Bogoliubov ansatz, we have shown that this mode can indeed be identified following a quench, and the roton gap energy can be measured from simple time-of-flight measurements. To ensure that this is not an artifact of our approximations, we have presented a fully self-consistent theory, which allows the condensate density to fluctuate and identified the parameter regimes where the non-self consistent results are expected to be qualitatively and quantitatively accurate. 
We predict that provided the inter-particle spacing $n^{-1/2}/l_{z} \leq 0.4$, the roton gap can be probed using a quench to within $10 \%$. While the quantitative magnitude of the effect we predict is contingent upon experimental realities, quench dynamics can complement other probes of excitations in dipolar gases \cite{blakie1, blakie2}. 

An important complication in most experiments is the presence of weak harmonic confinement. In a trap, the low lying collective modes are density oscillations, which occur on frequencies proportional to $\omega_{r}$, the radial trap frequency. The physics we describe here occurs on timescales $\tau_{\text{roton}} = 1/\Delta \sim$ms $\ll \tau_{\text{trap}} \sim 50$ms. On timescales much shorter than the trap period, the overall condensate density profile (which requires global mass transport) is relatively unaffected, and a local density approximation can be used. Using this approximation, we have verified that numerically that the oscillations in $n_{\text{ex}}$ persist for a few cycles in $\tau_{\text{roton}}$. On longer times, comparable to the trap period, collective density modes will wash out the oscillations in $n_{\text{ex}}$, and a better modeling of the condensate dynamics and the interaction between the condensed and non-condensed clouds is required. While such a theory can be performed, should experiments demand quantitative refinement, it is beyond the scope of the present paper. However, owing to the separation of timescales between $\tau_{\text{roton}}\ll \tau_{\text{trap}}$ in typical experiments, the oscillations we find here should be observable, without complications arising from trap effects.  Furthermore, nearly homogeneous Bose Einstein condensates have been experimentally realized by using box shaped traps \cite{zoranbox}.

\section{Summary and Conclusions}

In this work we have studied the non-equilibrium dynamics of a quasi-$2$D dipolar Bose gas, following a sudden quench in the interaction strength. Remarkably we find that quench dynamics can probe the excitation spectrum of the dipolar gas, which otherwise in equilibrium, displays \textit{no} pronounced spectral signature of the underlying roton mode. Physically, the reason for this is that on timescales $t \sim \tau_{\text{roton}} \ll \tau_{\text{trap}}, \tau_{\text{el}}$, excitations coherently oscillate in and out of the condensate. As roton excitations all oscillate at nearly the same frequency $\Delta$, this leads to a cooperative enhancement of the signal in correlation functions. In this respect, our bosonic system displays an interesting parallel with a Fermi gas with attractive interactions, where the quench to attractive interactions leads to an enhancement of the superconducting gap \cite{eliashberg, galitski, jacobsen, dobben, levitov}. It will be extremely interesting to study whether the \textit{non-equilibrium} enhancement of rotons could be a route to realizing (albeit metastable \cite{cooper, pedri}), quantum crystalline phases of matter.
 We hope that our work will not only motivate experiments to use quantum quenches to look for roton physics, but also further theoretical studies on the nature of spreading of correlations in dipolar systems. 

\section{Acknowledgements} It is a pleasure to thank Benjamin Lev, Kristian Baumann, Mingwu Lu and Rajdeep Sensarma for insightful discussions during the completion of the work, and Ryan Wilson for his careful reading of the manuscript. SN would like to thank V. Galitski for pointing him to Ref.~\cite{levitov}. SN is supported by JQI-NSF-PFC, AFOSR-MURI, and ARO-MURI.

\end{document}